\documentclass{WileyMSP-template}
\usepackage{xr-hyper}
\usepackage{color}
\usepackage{soul}
\usepackage{multicol}
\usepackage{amsmath}
\usepackage{cite}

\makeatletter
\newcommand*{\addFileDependency}[1]{
	\typeout{(#1)}
	\@addtofilelist{#1}
	\IfFileExists{#1}{}{\typeout{No file #1.}}
}
\makeatother
\newcommand*{\myexternaldocument}[1]{%
	\externaldocument{#1}%
	\addFileDependency{#1.tex}%
	\addFileDependency{#1.aux}%
}

\graphicspath{X:\paper
}
\myexternaldocument{SupplementaryInformation}
\begin{document}


\title{Electrically induced negative differential resistance states mediated by oxygen octahedra coupling  in manganites for neuronal dynamics}

\maketitle


\author{Azminul Jaman*}
\author{Lorenzo Fratino}
\author{Majid Ahmadi}
\author{Rodolfo Rocco}
\author{Bart J. Kooi}
\author{Marcelo Rozenberg}
\author{Tamalika Banerjee*}


\dedication{}

\begin{affiliations}
Azminul Jaman, Majid Ahmadi, Bart J. Kooi, Tamalika Banerjee \\
Zernike Institute for Advanced Materials, Groningen Cognitive Systems and Materials Centre, University of Groningen, Nijenborgh 4, Groningen, 9747 AG, The Netherlands.\\
Email Address: azminul.jaman@rug.nl, t.banerjee@rug.nl\\

Lorenzo Fratino, Rodolfo Rocco, Marcelo Rozenberg \\
CNRS Laboratoire de Physique des Solides, Université Paris-Saclay, 91405, Orsay, France

Lorenzo Fratino\\
CNRS Laboratoire de Physique Théorique et Modélisation, CY Cergy Paris Université, 95302 Cergy-Pontoise Cedex, France.\\

\end{affiliations}


\keywords{Negative differential resistance (NDR), Self-Oscillation, Leaky-integrate and fire(LIF) Neuron}

\begin{abstract}

The precipitous rise of consumer network applications reiterates the urgency
to redefine computing hardware with low power footprint. Neuromorphic computing utilizing correlated oxides offers an energy-efficient solution. By designing anisotropic functional properties in LSMO on a twinned LAO substrate and driving it out of thermodynamic equilibrium, we demonstrate two distinct negative differential resistance states in such volatile memristors. These were harnessed to exhibit oscillatory dynamics in LSMO at different frequencies and an artificial neuron with leaky integrate-and-fire dynamics. A material based modelling incorporating bond angle distortions in neighboring perovskites and capturing the inhomogeneity of domain distribution and propagation explains both the NDR regimes. Our findings establish LSMO as an important material for neuromorphic computing hardware.

\end{abstract}


\section{Introduction}
The high volume of consumer network, that precipitously rose, during the world-wide stay-at-home orders, few years ago, brought to light the urgent need to redefine computing hardware with low power footprint. In this, neuromorphic computing has gained steep interest due to the energy-efficient solution capabilities that new materials, such as oxides, offer for in-memory computing. For such non von Neumann computing approaches, correlated properties intrinsic to the oxides are commonly exploited. However, at the hardware level, these properties have been less explored for demonstrating the rich non-linear phenomenon present in biological systems. The emergence of collective order is the cornerstone of the class of correlated oxides \cite{dagotto2005open}, primarily stemming from strain tailoring at heterointerfaces and the intricate coupling between charge, spin, orbital, and lattice \cite{yamada2004engineered, sun1999thickness,li2000oxygen,rajeswari1998correlation,urushibara1995insulator}. Manganites with their intrinsic chemical inhomogeneities and complex phase diagrams are particularly relevant. Electronic transport in manganites is susceptible to external stimuli and leads to inhomogeneous spatial distribution of electronic phases with varying local properties \cite{burema2022temperature,dho2003oxygen,haghiri2003cmr,hwang1995lattice,van2022strain,van2024probing}. These coexisting and competing phases collectively result in non-linear responses of the order parameter, important for realizing nanoelectronic devices \cite{miao2020direct}. By driving such systems out of thermodynamic equilibrium, electrically or optically, such non-linear phenomena are realized, widening the repertoire of emergent properties. Here we design anisotropic functional properties in strained thin films of La$_{0.7}$Sr$_{0.3}$MnO$_{3}$ (LSMO), a well-established correlated oxide in the manganite family, by uniquely accommodating variations in the local octahedral tilts, across a twinned substrate of LaAlO$_{3}$ (LAO), with bond angle variations ranging from 180$^{\circ}$ to 150$^{\circ}$, at the adjacent atomic sites. Recent advancements in complex oxide engineering demonstrate an efficient control of oxygen octahedra coupling (OOC) at the film-substrate interface, at the atomic scale \cite{rondinelli2012control,liao2016controlled} , however their cooperative impact on the electronic properties remains unexplored.
We utilize this to demonstrate a non-linear response to current, in thin film devices of LSMO on LAO, driving the system to two distinct negative differential resistance (NDR) states, close to the metal-to-insulator (MIT) transition in LSMO, not reported for other types of volatile memristors. Electronic instabilities such as NDR are a key indicator of spatially distributed inhomogeneous phases, close to a phase transition, in a material \cite{krischer2010oscillations}. The electrical current ($I$) associated with these phase inhomogeneities manifest as a non-ohmic and negative slope for the electrode potential ($V$) \cite{li2019origin} and has been commonly observed in Mott insulators when driven out of thermodynamic equilibrium by electric field or current-induced Joule heating \cite{janod2015resistive,del2021spatiotemporal,kalcheim2020non}. We harness these NDR states to exhibit oscillatory dynamics in LSMO at different frequencies in a Pearson-Anson circuit with a capacitor and resistor in parallel and in series respectively, with the volatile memristor. We demonstrate the dynamic response of the entire volatile memristive network as a leaky-integrate-fire (LIF) neuron when the external stimuli are large enough to trigger an output, transiting the network to a low resistance state. Our material based modelling incorporates the distribution of Curie temperature ($T_{C}$) in the network, that corroborates with the bond angle distortions, as found from Scanning transmission electron microscopic (STEM) studies. The model captures the inhomogeneity of the domain distribution with increasing temperature that amplifies close to the phase transition temperature, where high resistance domains dominate. Joule heating driven temperature rise nucleates a high resistive barrier, that expands until it blocks charge migration and eventually snaps, equilibrating the sample and substrate temperature in the paramagnetic insulating state, explaining the observation of the low bias NDR. The unique observation of the second NDR is explained by the model, considering the enhancement of the barrier with increasing voltage, suppressing the ferromagnetic state and transiting to the paramagnetic metallic state, characterizing the phase transition beyond $T_{C}$ in LSMO. These findings, well explained by the rich correlated physics in LSMO widens the ambit of this material class for applications in low power computing, specifically within the realm of neuromorphic computing, opening new avenues towards more biologically plausible brain functionalities. 

\section{Results and Discussion}
\subsection{Strongly coupled phases and HAADF-STEM studies of the interface}
\begin{figure*}[h!]
    \centering
	\includegraphics[width=\linewidth]{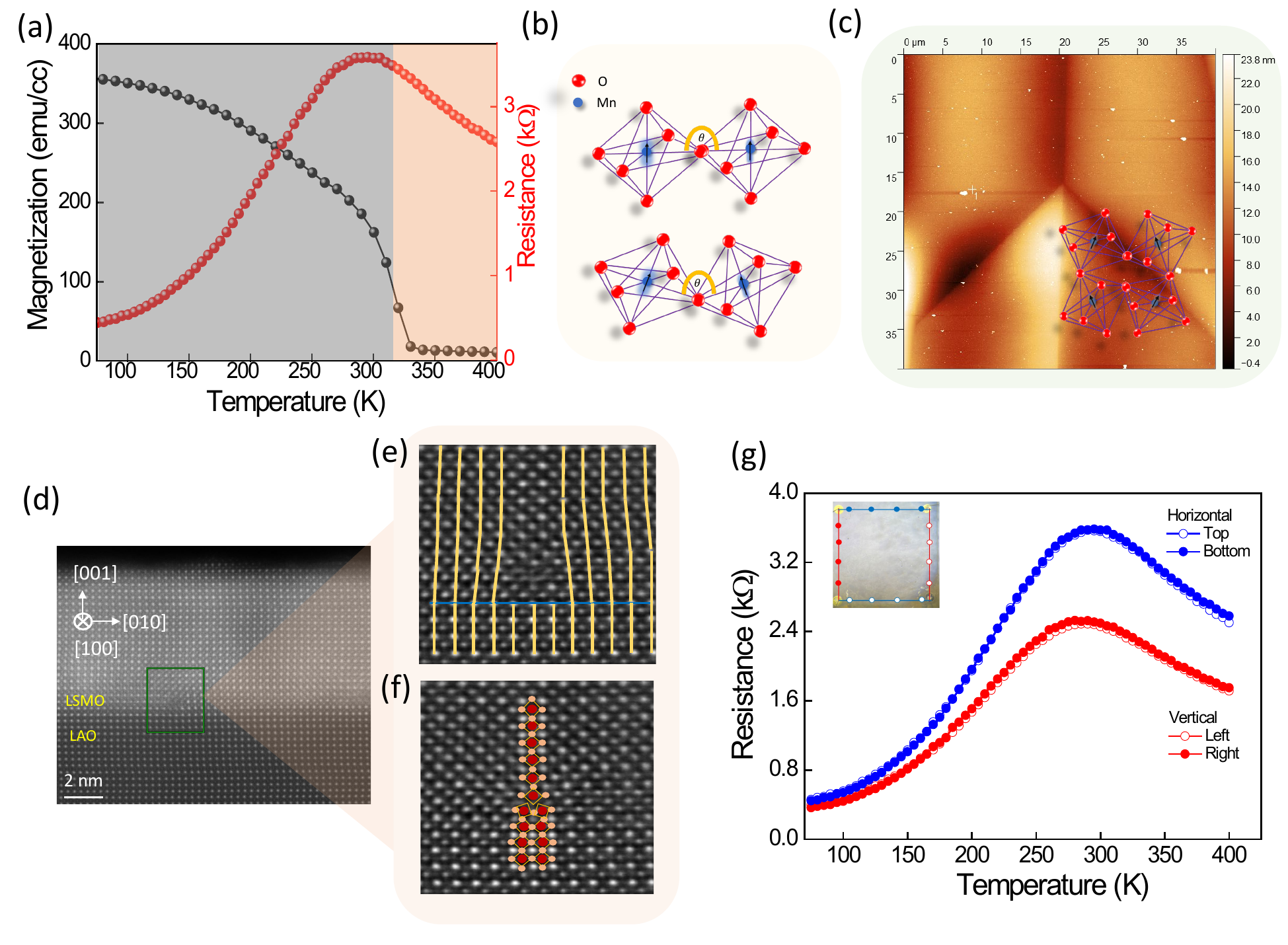}
	\caption{(a) shows the strong coupling between the electronic and magnetic phases. As the temperature increases the sample goes from ferromagnetic metal (light grey) to paramagnetic insulator (light orange) with a coinciding $T_{C}$ and $T_{MIT}$. (b) The octahedral rotations are schematically shown in the picture, the top panel shows perfectly parallel octahedra to each other enabling maximum probability of hopping of an electron, bottom panel shows distorted octahedra when hopping probability is reduced. (c) Atomic force microscope (AFM) image shows wedge disclinations and the inset shows the distorted arrangement of octahedra due to the occurrence of such wedge disclinations. (d) shows a HAADF-STEM image of a 10 nm thin LSMO film grown on a textured LAO substrate along the [100] zone axis. The green rectangle box encloses an area where a kink, representative of the twin domain boundary, is observed. (e) A zoomed-in view of the enclosed area is shown (iDPC); the yellow lines are drawn over the Sr atoms and the interface is represented by a vertical blue line. (f) Octahedral arrangements are shown in the LAO substrate and in the LSMO film. (g) Resistance is measured along two different directions with 1 \textmu A current, Blue filled (top) and open (bottom) circle represents resistance measured along horizontal direction, and red filled (left) and open circles (right) represent measured along vertical directions, shown in the inset.}
	\label{Fig_1}
\end{figure*}

    We characterize the coupled phase transition in a 10 nm strained thin LSMO film, on Lanthanum Aluminate (LAO) \cite{chen2020moire,jaman2023morphology}.  The double exchange interaction governs the functional properties in LSMO, where, in the parallel alignment, the adjacent oxygen octahedra ($MnO_{6}$) with the manganese (Mn) spins, allows the itinerant electron to hop from one $Mn^{3+}$ to the next $Mn^{4+}$ site. This alignment plays a crucial role in the strong coupling between electronic and magnetic phases, with a metal-to-insulator transition temperature $(T_{MIT})$ and Curie temperature $(T_{C})$ coinciding at 325 K shown in \textbf{Figure} \ref{Fig_1}a. Below 325 K, the film is a ferromagnetic metal and above 325 K, the film transforms into a paramagnetic insulating state. The conduction mechanism is hopping transport dominated where the probability of an electron moving from $Mn^{3+}$ to $Mn^{4+}$ depends on the bond angle between them shown in \textbf{Figure} \ref{Fig_1}b, top panel. This hopping probability is maximized if the bond angle between them is 180$^{\circ}$, and decreases with decreasing bond angle  (See, \textbf{Figure} \ref{Fig_1}b, bottom panel). A 10 nm thick LSMO film grown on a textured LAO substrate shows that the film is compressed along the in-plane direction and thus elongated in the out-of-plane direction, as revealed from reciprocal space map studies shown in \textbf{Figure} \ref{Figure_S2} (Supporting Information). Additionally, a slight smearing of the film peak has been observed, signifying an inhomogeneous strain in the film. The inhomogeneities in the strain arise from the structural twins boundaries present on the surface of LAO. When an epitaxial film of LSMO is deposited on such twinned substrates, the oxygen octahedron rotate and distort to preserve connectivity with the adjacent octahedra. \textbf{Figure} \ref{Fig_1}c shows an atomic force microscope picture of one such wedge disclination \cite{vermeulen2016unravelling,romanov2009application} and a schematic of the distorted octahedra due to such twins.

We employ high-angle annular dark field (HAADF) STEM to study the microstructure of the film and that of the interface with the substrate. In \textbf{Figure} \ref{Fig_1}d we show a cross-sectional TEM image of the film and the substrate. The green rectangular box highlights a small section of the interface between the film and the substrate, where we observe an irregular atomic arrangement within the LSMO correlated to the twin boundary in the LAO (not observable in this image). The zoomed image (see, \textbf{Figure} \ref{Fig_1}e) shows vertical yellow lines intersecting the interface over the La atoms and represents the substrate cations. Above the substrate, marked by the blue horizontal line, La/Sr cations of the LSMO film are mapped using the same trace of the vertical yellow lines. We observe that the columns of cations from the substrate to the film surface are distorted near and at the wedge disclinations. However, the cationic (La/Sr) columns are less distorted further from the kink on both sides. Focusing on the wedge disclination area shown in \textbf{Figure} \ref{Fig_1}f, we observe two parallel columns of $MnO_{6}$ oxygen octahedra, below the interface signifying an ordered arrangement of the perovskites. Once these two columns approach the interface, they merge into a single column in the film, signifying a missing column of cations and octahedra. The wedge disclination serves as nucleation points to maintain the geometrical constraint of the oxygen octahedral connectivity between the substrate and the film. To quantify the local strain distributions at and around the wedge disclination, we calculated the interatomic distances by analyzing the intensity profile of the cations both in the substrate and in the film and to form this defect that can be considered a wedge disclination shown in \textbf{Figure} \ref{Figure_S3} (Supporting Information). We characterize the electrical properties of the film by measuring the resistance ($R$) with temperature ($T$) along four different directions (see, inset \textbf{Figure} \ref{Fig_1}g). Overall the variation of the resistance vs temperature is similar for both directions with a larger resistance in the horizontal than in the vertical direction.\\

\subsection{Negative differential resistance regimes in voltage and current sweep}

\begin{figure*}[h!]
    \centering
	\includegraphics[width=\linewidth]{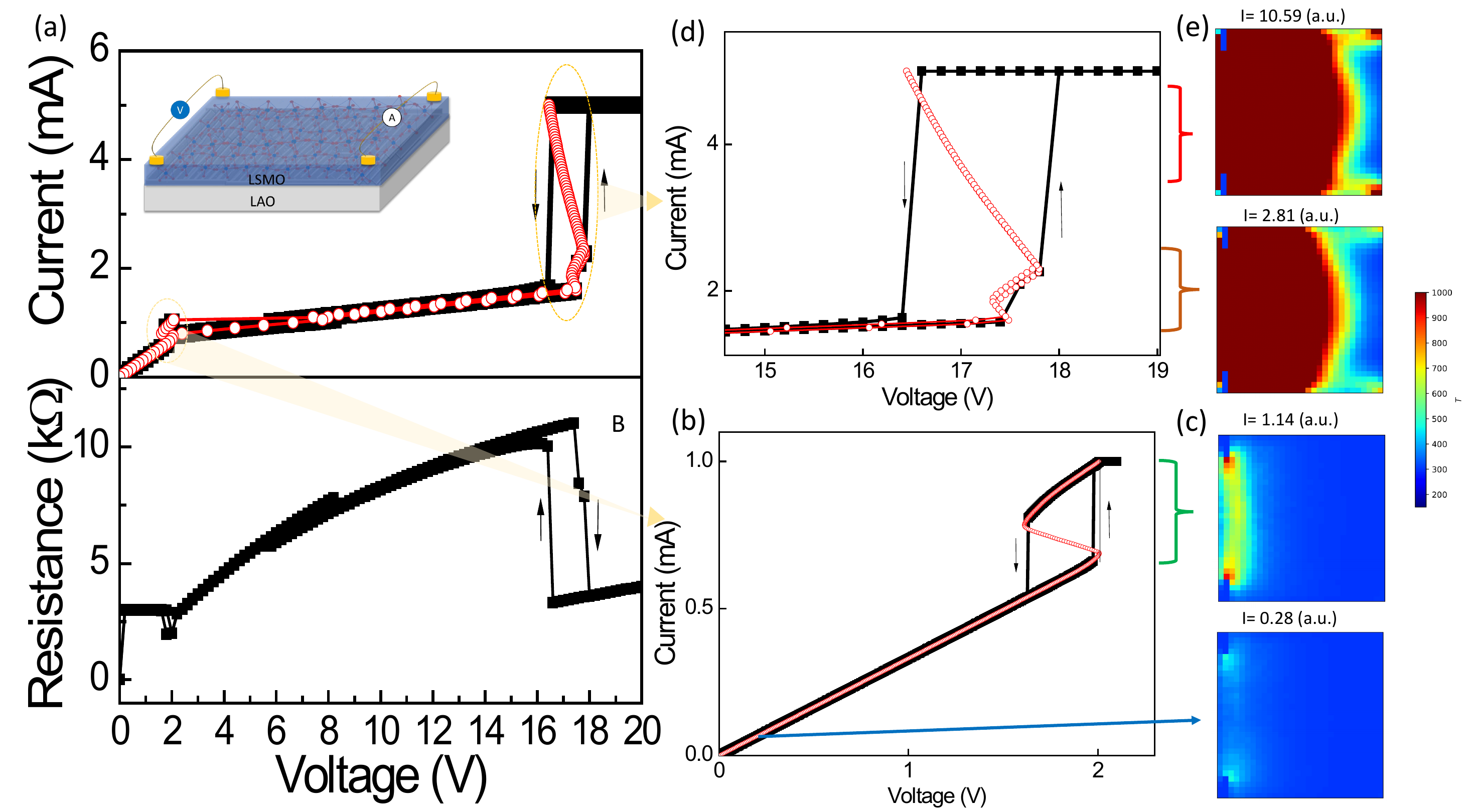}
	\caption{(a) Current (red circles) and Voltage (black squares) show linear and non-linear regimes, inset shows the schematic of the measurement geometry. The bottom panel is linked to the resistance calculated from the voltage sweeps (b) Voltage sweep at low bias with more data points and (c) corresponding simulated temperature evolution at two different voltages.(d) A zoom into the high bias non-linear regime reveals that the simulated (e) temperature increases abnormally and is concentrated between the two electrodes. As a result of Joule heating for both the bias regimes the current increases abruptly, whereas in the current sweep, a sharp onset to S-type NDR is observed.}
	\label{Fig_2}
\end{figure*} 
Resistive switching is studied by sweeping either the voltage ($V$) or the current ($I$) at a fixed substrate temperature of 300 K. Two distinct regimes at different voltage bias are found in these volatile memristors, as shown in \textbf{Figure} \ref{Fig_2}a. Black squares represent the voltage sweep, and the red open circles represent the current sweep. In the first regime, close to 2 V, a small jump in the current is observed, the corresponding resistance drop is shown in the plot below. In the second regime, the resistance (current) increases (decreases) upto $\sim$ 16 V, beyond which the resistance abruptly decreases shown in \textbf{Figure} \ref{Fig_2}a, bottom panel. Looking closely at the small but abrupt transition region at $\sim$ 2 V by collecting more data points, we find a sharp hysteretic increase in current with voltage, while in the current sweep, a non-linear snap back in the sensed voltage is recorded, corresponding to a negative differential resistance (NDR) behavior illustrates in \textbf{Figure} \ref{Fig_2}b. A distinct second NDR regime is observed at a higher voltage of ($\sim$ 17.4 V) (see, \textbf{Figure} \ref{Fig_2}d). Multiple NDRs as observed in our LSMO films on LAO have not been reported either on STO \cite{salev2021transverse,salev2024local} or in Mott insulators \cite{gibson2016accurate,kumar2017chaotic,nath2023thermal,das2023physical} and presents a new opportunity for studying oscillatory dynamics. Mott resistor network simulations are performed and reveal the dynamic evolution of the Joule heating induced local hot spots, underneath the sourcing contacts that enhances further with increasing voltage bias and is responsible for the NDR regimes as shown in\textbf{Figure} \ref{Fig_2}c and \textbf{Figure} \ref{Fig_2}e. At low voltage bias (shown for two different voltages), the temperature rises slowly corresponding to the ohmic region until it reaches the critical phase transition temperature ($T_{C}$). The resistance collapses at $T_{C}$ and the ferromagnetic phase transits to a paramagnetic insulating state (\textbf{Figure} \ref{Fig_2}b and \textbf{Figure} \ref{Fig_2}c), manifested by the first NDR. At higher bias, the temperature rises significantly ($\sim$ 1000 K) between the electrodes, leading to an inhomogeneous and uneven temperature distribution across the entire film with the onset of the second NDR regime (\textbf{Figure} \ref{Fig_2}d and \textbf{Figure} \ref{Fig_2}e).

\subsection{Simulated and experimental results}
\begin{figure*}[h!]
    \centering
	\includegraphics[width=\linewidth]{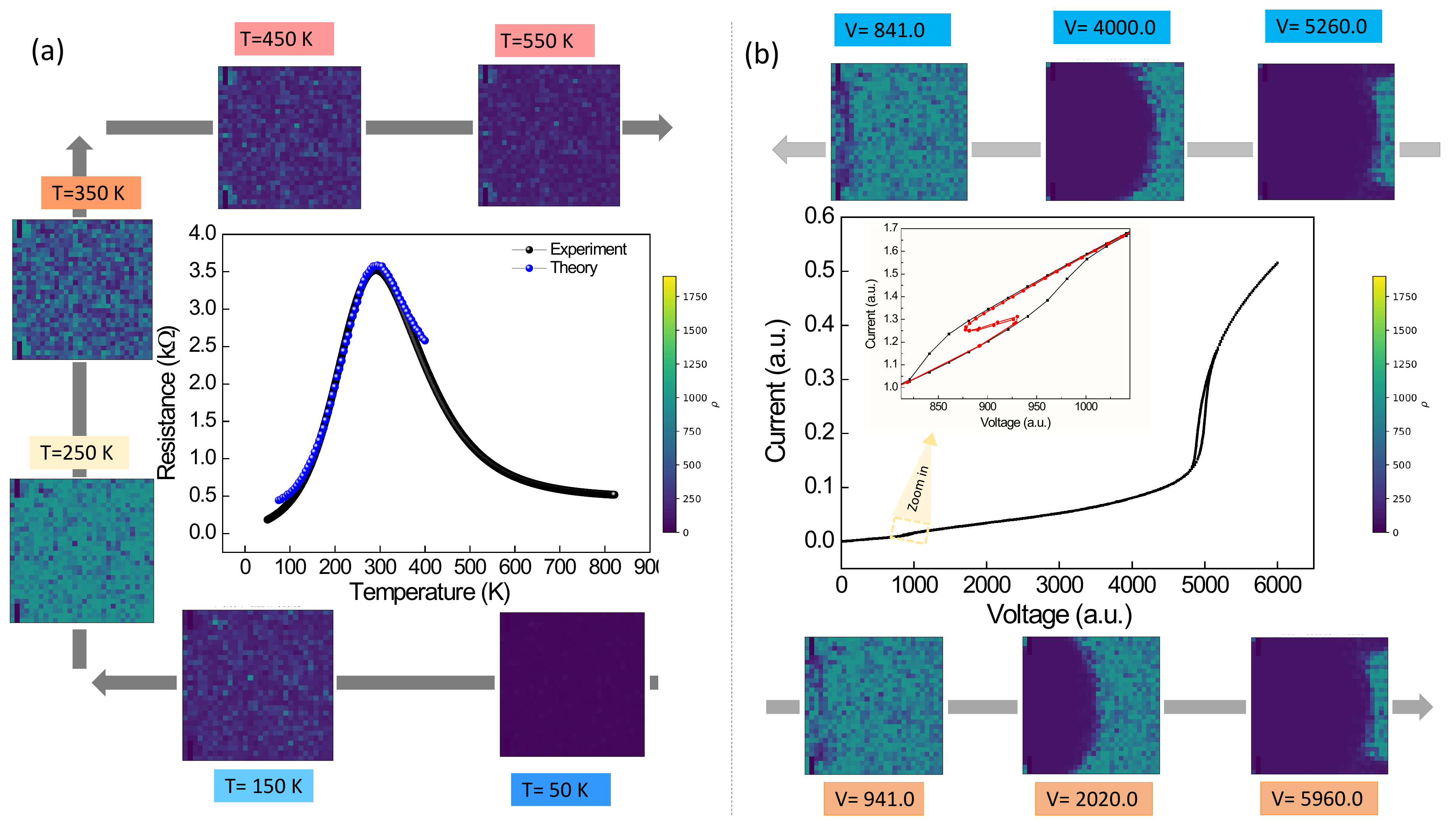}
	\caption{(a) Experimental (blue-filled circles) and simulated (black-filled circles) resistance vs temperature curves along with the simulated resistor network images of low and high resistive states at different base temperatures: 50 K, 150 K, 250 K, 350 K and 450 K. (b) Simulated current/voltage characteristics qualitatively captures small bias NDR (inset) similar to Fig 2B and high bias hysteretic threshold switching. The resistor network map with increasing (bottom panel) and decreasing (top panel) voltages is shown.}
	\label{Fig_3}
\end{figure*}

We also perform resistor network simulations \cite{quintero2007mechanism,rozenberg2010mechanism} as shown in \textbf{Figure} \ref{Fig_3}a. The experimental limitations on recording the resistance changes with temperature beyond 400 K are eased by the realistic approach taken in the theory as described in section S\ref{S1} (Supporting Information). The correspondingresistor maps have been shown at a different fixed substrate temperature. At low temperatures of $\sim$ 50 K, the LSMO film is in a ferromagnetic metallic state with low resistive domains that are homogeneously distributed. With increasing temperature, we find an inhomogeneity in the distribution of domains with the formation of high resistive domains at 150 K. At 250 K, close to $T_{C}$ , the size of the high resistive domains further increase along with an overall increase in the resistance of the film. As the transition temperature ($\sim$ 350 K) is reached, most of the low-resistive domains transform into high-resistive domains. Interestingly, we find that beyond the transition temperature, at 450 K, the low resistive domains start to form again, enhancing with further increase of the temperature as manifested in the overall lowering of the resistance shown in the $R$-$T$ plot. 

A qualitative agreement between experimental and simulated I-V is achieved (see, \textbf{Figure} \ref{Fig_3}b). The resistor network I-V sweep captures both the low and high bias switching and the NDR regimes. At small voltages and low temperatures, we observe an ohmic relation, where the charges are all homogeneously
distributed. A further increase in the voltage results in an increase in the temperature of the film through Joule heating, leading to an increase in its resistance. Once the local temperature starts approaching the
critical temperature $(T_C)$, a barrier of high resistive state starts to form around the electrodes. Such a barrier behaves as a bottleneck for the charge migration resulting in an inhomogeneous distribution of charges and causing the temperature to rise drastically in this region. The barrier gradually grows and eventually snaps the path between the electrodes leading to a drop in the sample temperature and finally equilibrates the sample and substrate temperature. By further increasing the voltage, we observe the low bias NDR region visible in our simulation (see, \textbf{Figure} \ref{Fig_3}b, V=1500 a.u.) and experiment (see, \textbf{Figure} \ref{Fig_2}a, V= 2 V), where the current drops. The width of the barrier increases in such a way that the current diminishes while a column begins to become more insulating until it completely switches to a higher resistive state. Once the barrier covers the entire sample, the homogeneity of charge distribution is reestablished, thus showing an ohmic behavior in a paramagnetic insulating state (see, Figure 3b, between
$\sim$ 2000 a.u. to $\sim$ 4500 a.u.). At higher voltages, the width of the filament formed covers more regions of the network, as shown in see, \textbf{Figure} \ref{Fig_3}b. Once the filament formed in the paramagnetic state covers more than half of the sample, the sample temperature on the right edges rises abruptly due to the short resistive path. This creates two new filaments transverse to the first one illustrates in \textbf{Figure} \ref{Figure_S5} (Supporting Information), that eventually leads to the jump in current, similar to the experimental observations.
\

\subsection{Oscillatory dynamics of an artificial neuron}
\begin{figure*}[h!]
    \centering
	\includegraphics[width=\linewidth]{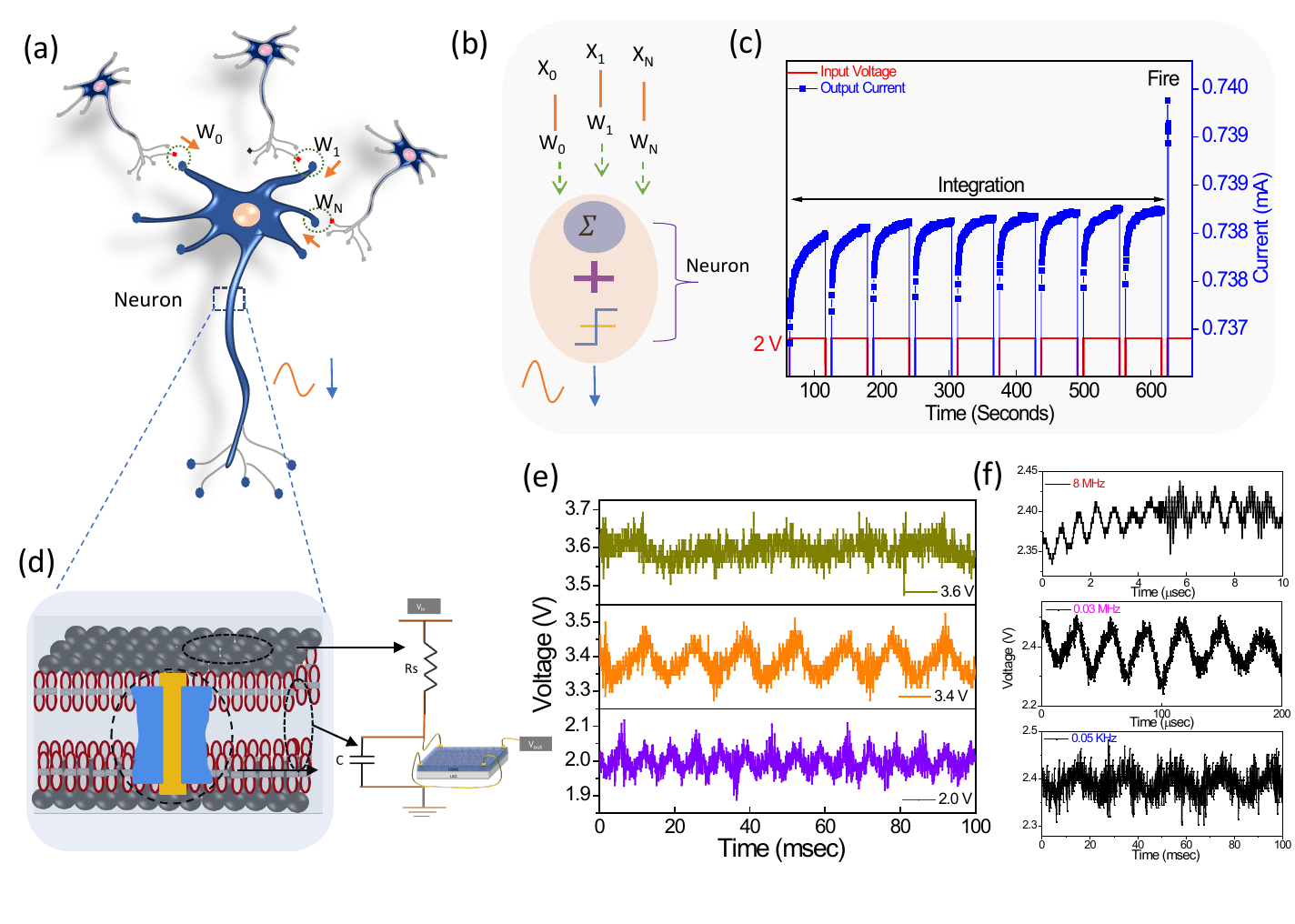}
	\caption{(a) A schematic representation of multiple neurons connected to a single neuron (b) Mathematical operation that explains the signal propagation in the human brain. Neurons being the central part of signal propagation is characterized by mainly two operations that consist of integration of the incoming signals and activation. (c) The demonstration of leaky-integrate-fire neuronic behavior at constant 2 V square pulses with a delay time of 5 seconds.(d) Zooming into a single neuron shows an operation of signal propagation through the membrane and its equivalent electrical circuit. (e) Shows a voltage-tunable oscillator whose maximum oscillation frequency is harnessed (f) by changing the electrical time constant (RC) varying the capacitor (100 µf, 10 nf, and 100 pf ). Oscillation frequencies in the MHz range is observed.}
	\label{Fig_4}
\end{figure*}

      Physical processes leading to instability are relevant for oscillatory computing processes inspired by the brain architecture. In the brain, a single neuron is connected to multiple other neurons via synapses, that feeds the signal, typically charges, for integration in the cell membrane over time \cite{yi2018biological}. Once the accumulated charges reach a threshold, a spike is emitted by opening an ion channel that propagates to the next neuron as shown in \textbf{Figure} \ref{Fig_4}a. Mathematically, a combination of an integration and activation function can mimic a single neuron (see, \textbf{Figure} \ref{Fig_4}b). Here we show how Joule heating induced switching in LSMO has been utilized to demonstrate such an artificial leaky, integrate and fire neuron (LIF). We exploit the low bias NDR voltage regime and the volatile nature of the switching and subject the film to a train of pulses of amplitude 2 V with a time delay of 5 seconds in between each pulses. A critical point is attained when the remnant heat due to the application of successive pulsed voltages in combination with the delay time leads to a sharp jump in current and establishes LIF functionalities (see, \textbf{Figure} \ref{Fig_4}c). A schematic anatomy of a biological neuron and its electronic analog is shown in \textbf{Figure} \ref{Fig_4}d, where the LSMO film act as the ion channel as in the biological neuron. The presence of multiple NDRs that we uniquely find in our LSMO film further led us to establish its potential as an oscillatory neuron, when connected to a circuit described above. We have used a simple $RC$ circuit illustrates in \textbf{Figure} \ref{Fig_4}d, such as the Pearson-Anson circuit \cite{salverda2022epitaxial,hamming2024mixed}, to establish LSMO as a self-oscillator. In this circuit, a DC voltage was fed as an input and the resulting Joule heating led to a change in the resistance of the film to a high resistance state. Once the film transforms to the high resistive state, part of the current starts charging the capacitor until it is almost fully charged, meanwhile lesser power across the film allows it to cool down. This leads to an oscillation of the output voltage between high and low values, whose frequency is found to be dependent on the magnitude of the DC input voltage as shown in \textbf{Figure} \ref{Fig_4}e. As we increase the input voltage from 1.95 V to 3.4 V, the oscillation frequency decreases, demonstrating a voltage-tunable oscillator. To harness the maximum oscillation frequency, we have varied the external electrical time constant (RC) from millisecond (ms) to picosecond (ps), resulting in an increased oscillating frequency from kilohertz (kHz) to sub-megahertz (MHz) range
 
\section{Conclusion}
Exploiting the intrinsic strongly coupled electronic and magnetic phase transition in LSMO, along with the design of the local octahedral distortion by using a twinned substrate of LAO, we uniquely demonstrate the presence of at least two NDR regimes when the film is driven out of thermodynamic equilibrium by electrically triggering it. Whereas the first unstable low bias regime, the first NDR regime, is due to inherent ferromagnetic-metallic to paramagnetic insulating transition (25), the second NDR at high bias is due to a filamentary formation promoted by the effective short-circuit in the measurement of the non-local current. Mott resistor network simulation shows that our experimental findings agree well with the theory for an inhomogeneous distribution of $T_{C}$’s arising due to local distortions of the oxygen octahedra revealed by STEM studies. The multiple NDR states is used to establish a voltage-tunable oscillator that exhibit periodic oscillation of these resistive states. The interplay between intrinsic thermal switching time and external electrical time ($RC$) constant controls the system dynamics to generate signals ranging from sub-kHz to MHz frequency range. With the growing need for non-linear electronic transport-based system for spiking neuronal behavior in brain-inspired computing, the collective dynamics intrinsic to LSMO, yielding such diverse functionalities lends itself as an important material for biologically plausible brain functionalities.


\section{Experimental Section}
 
 Sample growth and fabrication: LSMO films of thickness 10 nm are grown on LAO (001) substrate using pulsed laser deposition (PLD). The 248 nm pulsed KrF laser is used to ablate LSMO target at a fluence of 2 J/cm$^{2}$ and repetition rate of 1 Hz. The growth is done at a constant oxygen pressure of 0.35 mbar and the substrate temperature is kept constant at 750 $^{\circ}$C. After the growth, the samples are cooled down to room temperature at a rate of 10 $^{\circ}$/C in the presence of 100 mbar of $O_{2}$. 10 nm Titanium (Ti) with a capping of 50 nm of Gold (Au) is deposited at the four corners of the sample to make van der Pauw contacts for electrical measurements using optical UV lithography.\\

 Electrical characterization: Keithley 2410 source meter is used to source a DC current varying in magnitude from 1 uA to 7 mA and the corresponding voltage drop is measured in a four-wire configuration by sweeping the sample temperature to obtain the sheet resistance ($R$) dependence on temperature ($T$). Current/voltage sweep is done either by sweeping current or voltage at different sample temperatures. Under a constant bias, voltage oscillations are recorded using a Keysight DSOX1204A oscilloscope.\\

 Scanning transmission electron microscopy (STEM): Thin 20 $\mu$m X 20 $\mu$m long lamella is prepared using Helios G4 CX dual beam system with Ga focused ion beam from a 5 mm X 5 mm LSMO sample for STEM imaging. HAADF-STEM and iDPC-STEM are both used to detect heavy cations, as well as lighter oxygen ions. We image on multiple regions of the sample to obtain convincing statistics. The raw data is interpreted using Thermofisher Velox software\\

\medskip
\textbf{Supporting Information} \par 
Supporting Information is available from the Wiley Online Library or from the author.

\medskip
\textbf{Acknowledgements} \par 
A. Jaman acknowledges financial support from CogniGron Centre. Technical support from J. G.
Holstein, H. H. de Vries and F. H. van der Velde are acknowledged. A.J. and T.B. acknowledge Konstantinos Panagiotis Rompotis, Walter Quiñonez, Job J L van Rijn and Anouk Goossens for
scientific discussions. This work was realized using NanoLab-NL facilities.

\medskip
\textbf{Conflict of Interest}

The authors declare no conflicts of interest.

\medskip

\textbf{Data Availability Statement}

The data that support the findings of this study are available from the
corresponding author upon reasonable request

\medskip

%
\bibliographystyle{MSP}
\bibliography{Bibliography}

\section{Table of Contents}

We demonstrate various neuronal functionalities such as leaky-integrate, fire (LIF) and oscillatory neurons in a single thin film of  La$_{0.7}$Sr$_{0.3}$MnO$_{3}$. STEM shows that the modification in Oxygen octahedral connectivity leads to various co-existing phases captured by Mott resistor network simulation. We have used the co-existing phases to showcase diverse brain functionalities using a  $RC$ circuit.

\begin{figure}[h]
\medskip
  \includegraphics[width=\linewidth]{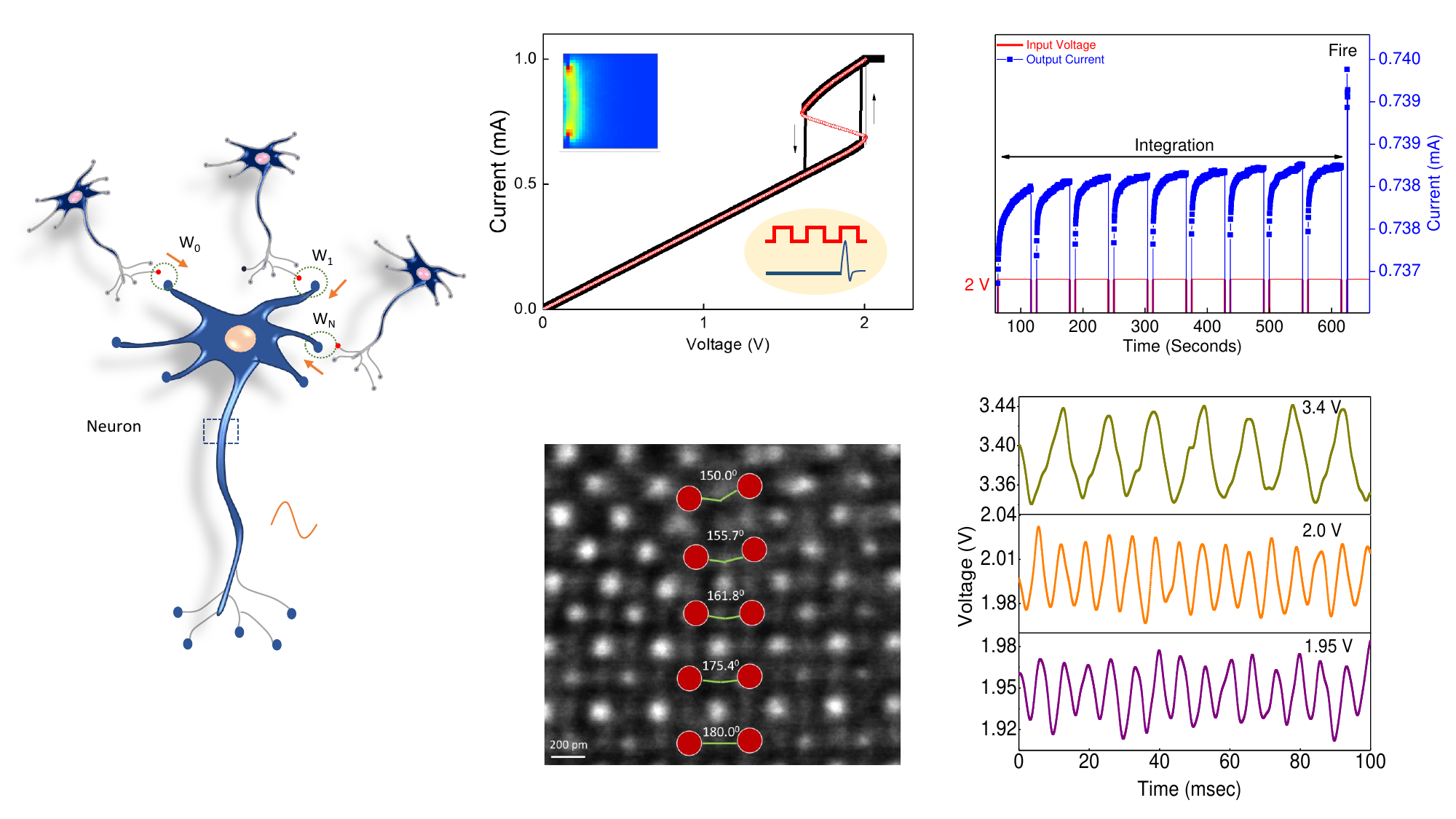}
  \medskip
\end{figure}

\end{document}



\newcommand*{\myexternaldocument}[1]{%
	\externaldocument{#1}%
	\addFileDependency{#1.tex}%
	\addFileDependency{#1.aux}%
}

 \newcommand{\beginsupplement}{%
         \setcounter{table}{0}
         \renewcommand{\thetable}{S\arabic{table}}%
         \setcounter{figure}{0}
         \renewcommand{\thefigure}{S\arabic{figure}}
         \setcounter{equation}{0}
         \renewcommand{\theequation}{S\arabic{equation}}%
      }

\graphicspath{X:\paper}
\beginsupplement

\title{Electrically induced negative differential resistance states mediated by oxygen octahedra coupling  in manganites for neuronal dynamics}
\maketitle

\author{Azminul Jaman}
\author{Lorenzo Fratino}
\author{Majid Ahmadi}
\author{Rodolfo Rocco}
\author{Bart J. Kooi}
\author{Marcelo Rozenberg}
\author{Tamalika Banerjee}

\begin{affiliations}
Azminul Jaman, Majid Ahmadi, Bart J. Kooi, Tamalika Banerjee \\
Zernike Institute for Advanced Materials, Groningen Cognitive Systems and Materials Centre, University of Groningen, Nijenborgh 4, Groningen, 9747 AG, The Netherlands.\\
Email Address: azminul.jaman@rug.nl, t.banerjee@rug.nl

Lorenzo Fratino, Rodolfo Rocco, Marcelo Rozenberg  \\
CNRS Laboratoire de Physique des Solides, Université Paris-Saclay, 91405, Orsay, France

Lorenzo Fratino\\
CNRS Laboratoire de Physique Théorique et Modélisation, CY Cergy Paris Université, 95302 Cergy-Pontoise Cedex, France.

\end{affiliations}

\section{S1: Mott resistor network modelling}
\label{S1}

\begin{figure*}[h!]
    \centering
	\includegraphics[width=0.6\linewidth]{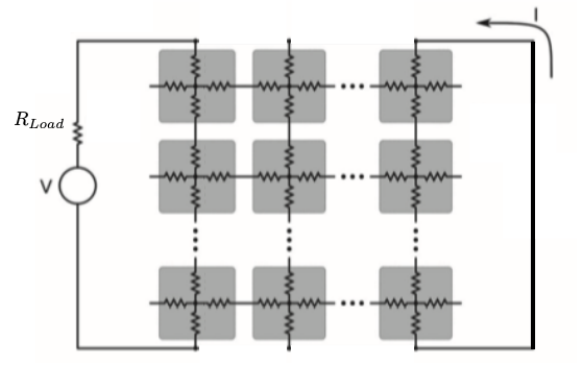}
	\caption{Sketch representation of the Mott resistor network model and simulation circuit. A bias voltage is inserted on the left side of the network while a current is measured on the right side}
	\label{Figure_S1}
\end{figure*} 
We model and numerically solve our system as the Mott resistor network model as schematized in (Fig. \ref{Figure_S1}). By applying $V_{App}$ a voltage to this circuit, the load resistance $R_{load}$ makes a voltage divisor circuit. Each site of our grid of size L $\times$ W (with L = 30 and W = 30) is represented by 4 resistors with an assigned resistivity given by the temperature by following this phenomenological curve

\begin{equation} 
 R(T)= \frac{R_{0}e^\alpha}{m}+ \frac{m-1}{m}  \frac{R_{1}e^{\frac{(\gamma-\lambda)T}{T_{C}}-\lambda}}{(1+e^{(-\lambda*(T-T_{C}))})}
\end{equation}
The values used in the simulations are: m = 3, $\lambda$ = 8.12 A.U., $T_{C}$ = 280 K, $R_{0}$ = 0.05 A. U., $R_{1}$ = 0.4 A.U., $\gamma$ = 4.64 and

\begin{equation}
    \alpha= 
\begin{cases}
    \frac{4.93}{Tc}T& \text{for} \quad T > Tc\\
    4.93& \text{for} \quad T< Tc\\
\end{cases}
\end{equation}

In order to simulate domains structures on the film we generate a Gaussian random grid of critical temperatures $T_{C}$ with mean value $310$ K and standard deviation $60$ K. This means that at the same temperature, each element of our network presents a different value of resistivity. Finally, we evaluate the voltage of the sample $V_{S}$

\begin{equation} 
 V_S= \frac{R_{S}}{R_{S}+R_{L}}V_{APP}
\end{equation}

where $R_{S}$ denoted the total resistance of the resistors network. As done in {\cite{salev2021transverse}, the local temperature of each site is updated at each numerical simulation "sweep" with the following formula (thermal diffusion):

\begin{equation} 
\od{T_{ij}}{t}=\frac{V^{2}_{ij}}{C_{v}R_{ij}}+\frac{k_{h}}{C_{v}}(5T_{ij}-T_{subs}- \sum_{<ik>}^{first\ neighbours} T_{ik})
\end{equation}

Where the indices $i$ and $j$ denotes the coordinates of the site in the network,
$C_{v}$ = 1.0 the specific heat of the system, $k_{h}$ = 0.01 the thermal conductivity
and $\sum_{<ik>}^{first\ neighbours}$ denotes the summation over the first neighbouring sites around the one with coordinates $ij$. In order to address a more realistic topology of the clusters in comparison to the experiment, the electrodes are put on the left side of the sample and their dimension is of width 1 and length of 3.

\pagebreak
\section{S2: Growth and structural characterization}

\begin{figure*}[h!]
    \centering
	\includegraphics[width=1\linewidth]{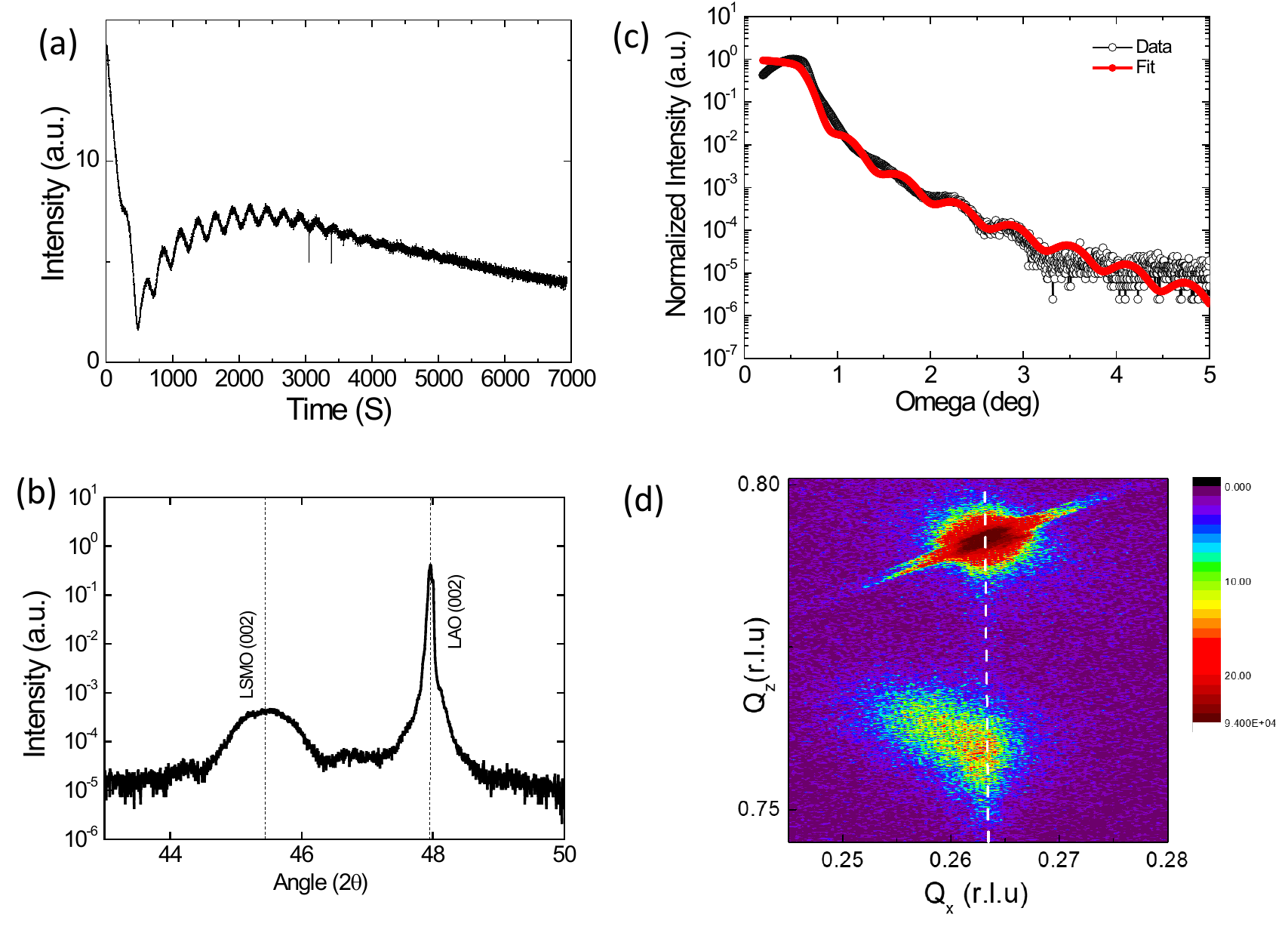}
	\caption{(a) Reflection high energy electron diffraction (RHEED) shows intensity oscillation signifying,(2D) layer-by-layer epitaxial growth of the film. (b) X-ray diffraction pattern of phase pure LSMO on the LAO substrate was observed. (c) X-ray reflectivity verifies that the thickness is about 10 nm. (d) reciprocal space map at (103) peak shows that the film is strained in the plane}
	\label{Figure_S2}
\end{figure*} 

\pagebreak
\section{S3: Strain inhomogeneities due to the wedge disclination on top of the twin}

\begin{figure*}[h!]
    \centering
	\includegraphics[width=1\linewidth]{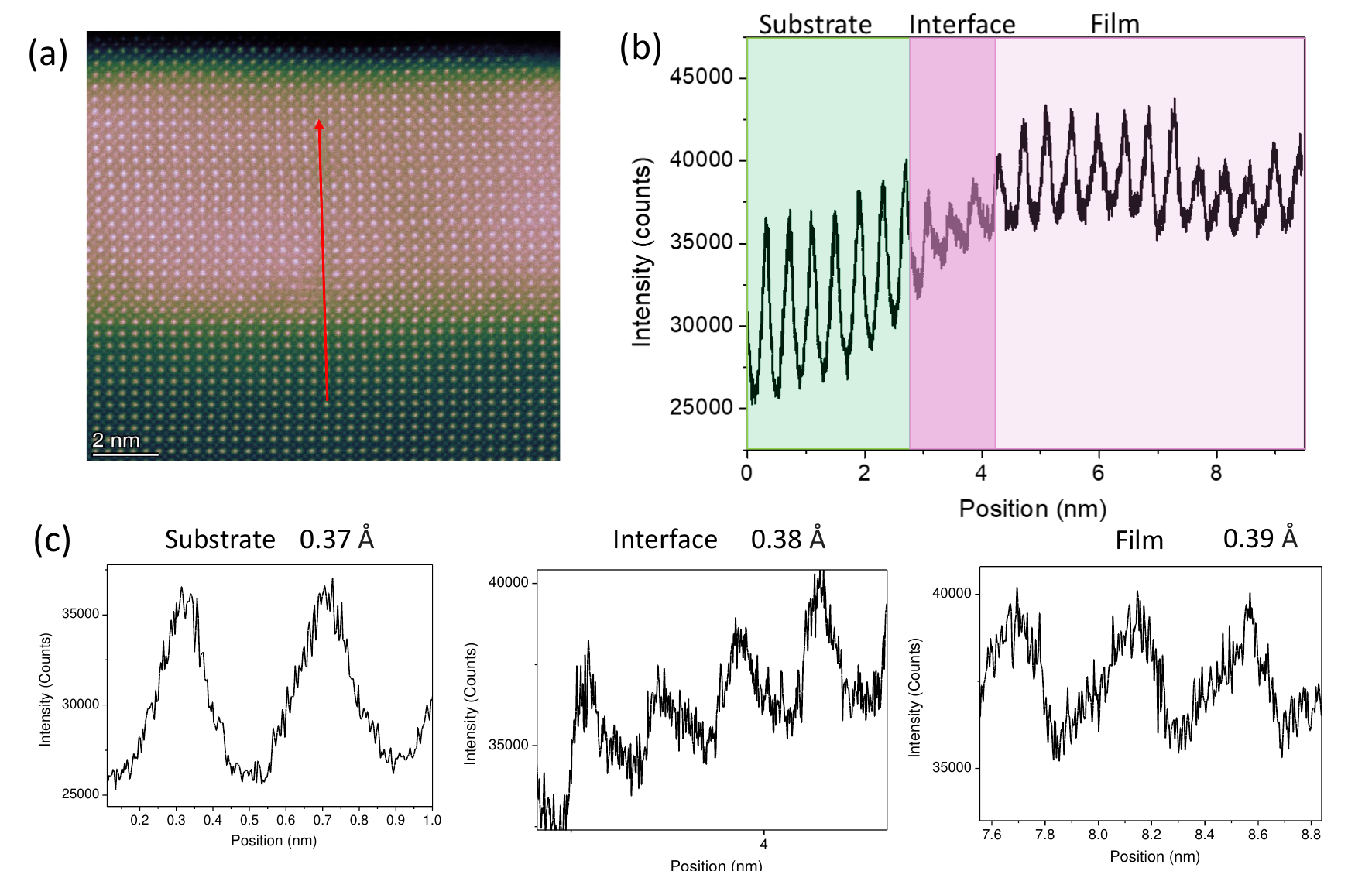}
	\caption{(a) HAADF-STEM images with two distinctive colours show the cationic arrangement between the film and the substrate. A line (red) has been drawn along the twin boundary from the substrate to the film. (b) The corresponding intensity profile is shown where the substrate region is marked by light green, the film is marked by light pink and the interface is marked by deep pink. (c) The interatomic distances were calculated using peak-to-peak distance, which is found to be 0.37 Å  for the substrate, 0.38 Å for the film at the interface and 0.39 Å for the film near the surface. It can be noted that the film is more elongated at the out-of-plane direction at the surface than at the interface.}
	\label{Figure_S3}
\end{figure*} 

\pagebreak
\section{S4: Intensity distributions and Bond angle modifications }

\begin{figure*}[h!]
    \centering
	\includegraphics[width=1\linewidth]{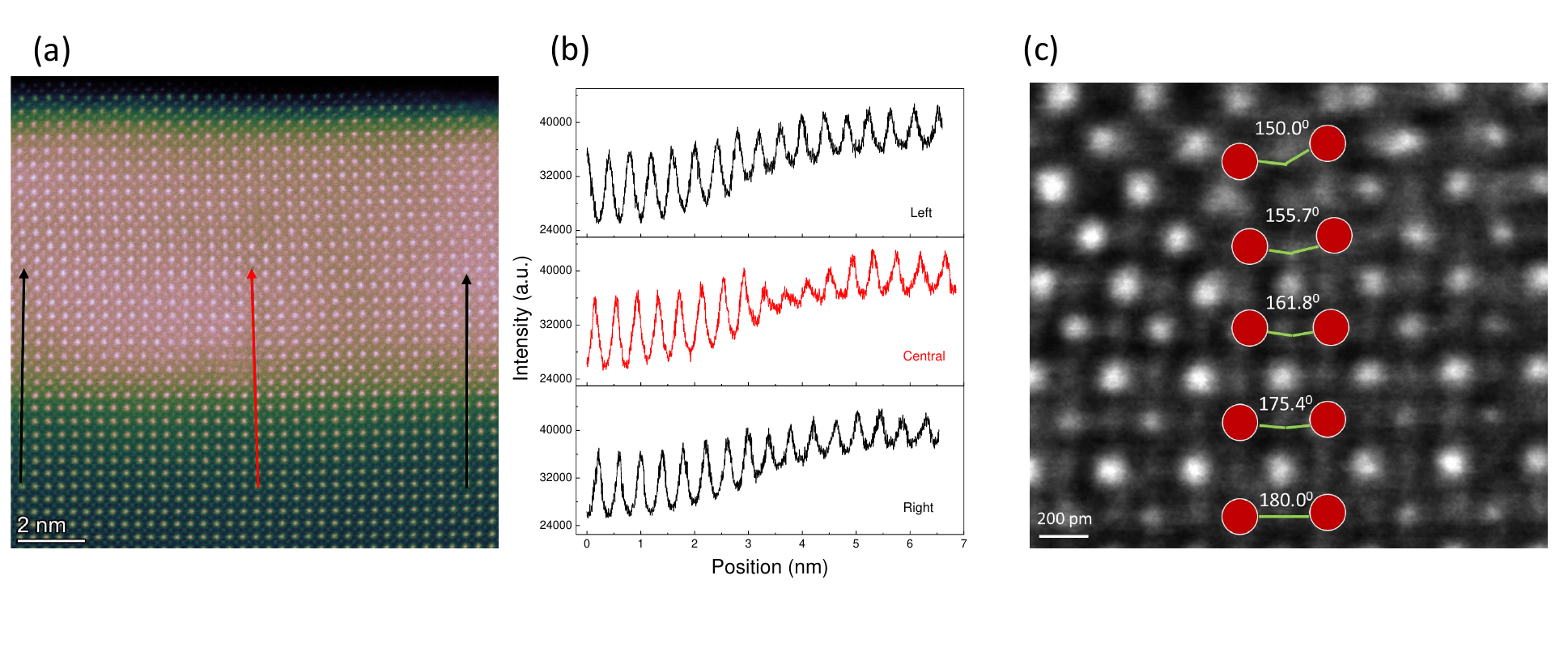}
	\caption{(a) To measure the intensity profile on both sides of the twin boundaries three lines of same length have been drawn through the wedge disclination on top of the twin (red arrow), on both sides (black arrows) of the twin. (b) We can see that for both the black lines on either side of the red line, the intensity of the cations from the substrate to the film remain the same with almost no loss of the intensity at the interface. The intensity is decreased through the twins at the interface. (c) iDPC-HAADF image shows how the bond angles have been changed between two adjacent Mn (red spheres) atoms at the interface}
	\label{Figure_S4}
\end{figure*} 

\pagebreak
\section{S5: Equivalent circuit of the four terminal measurement geometry of the sample, R-T at different current bias and I-V at different temperatures }

\begin{figure*}[h!]
    \centering
	\includegraphics[width=1\linewidth]{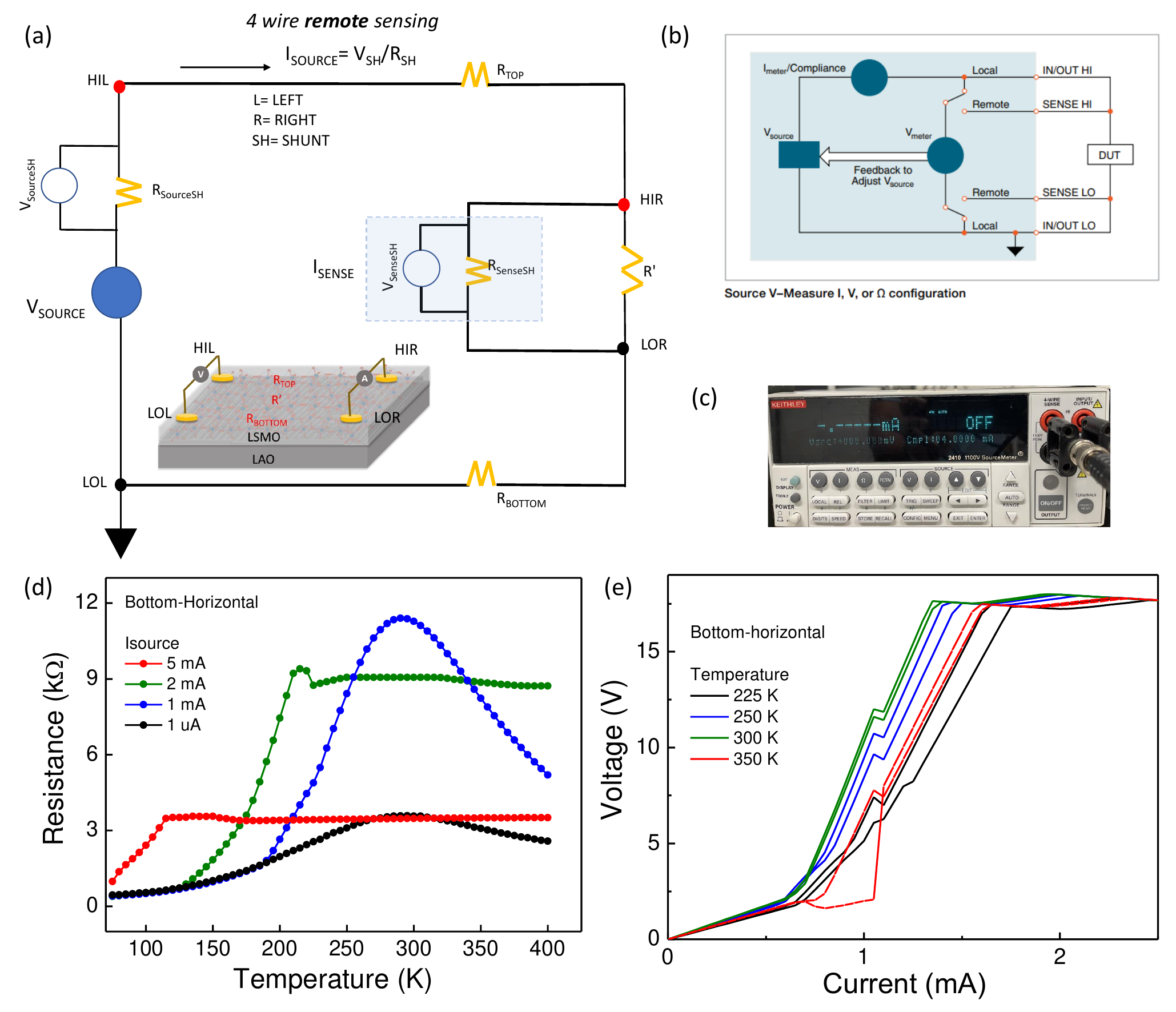}
	\caption{(a) Equivalent measurement circuit of the 4 wire sensing of the sample where RSourceSH, RTOP, RBOTTOM and R' are the resistances in the sourcing path, near the top right contact, near the bottom right contact and the remainder of the sample respectively. (inset shows the sample with contacts). (b) Keithley 2410 in voltage source (Vsource) and Current sense mode (Isense) (Keithley 2400 Datasheet) (c) Shows the Keithley source meter in the same configuration;  we measure current while sourcing voltage. (D) Resistance vs temperature measured at different bias currents. We observe the $T_{C}$’s at 1 µA and 1 mA to be the same whereas they gradually shift to a lower value with increasing current bias with an lowering of the overall resistance. (E) Voltage vs current sweep at different temperatures shows the resistive switching as well as the NDR. Both the measurements agree well, showing the resistance lowering with increasing bias current, eventually leading to unstable NDR regimes.}
	\label{Figure_S5}
\end{figure*} 

\pagebreak
\section{S6: Oscillation in each of the four directions of the film }

\begin{figure*}[h!]
    \centering
	\includegraphics[width=1\linewidth]{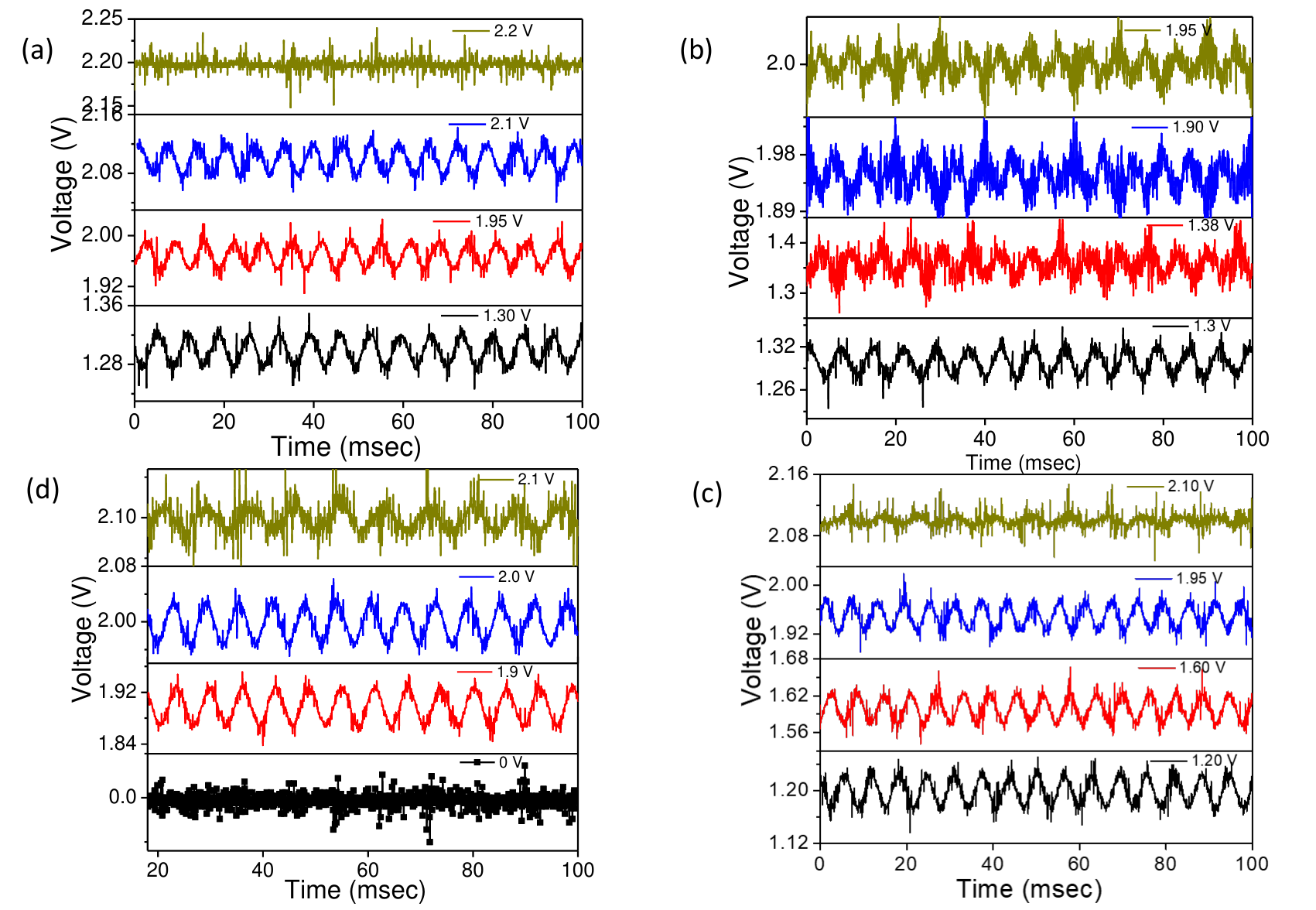}
	\caption{Voltage oscillation with a series resistance $R_{S}$ of 220 $\Omega$ and capacitance C of 100 µF at room temperature in  (a)Upper horizontal (b) Bottom horizontal (c) Right vertical (d) Left vertical}
	\label{Fig_S6}
\end{figure*}

\pagebreak
\section{S7: Oscillation metrices}
\begin{figure*}[h!]
    \centering
	\includegraphics[width=1\linewidth]{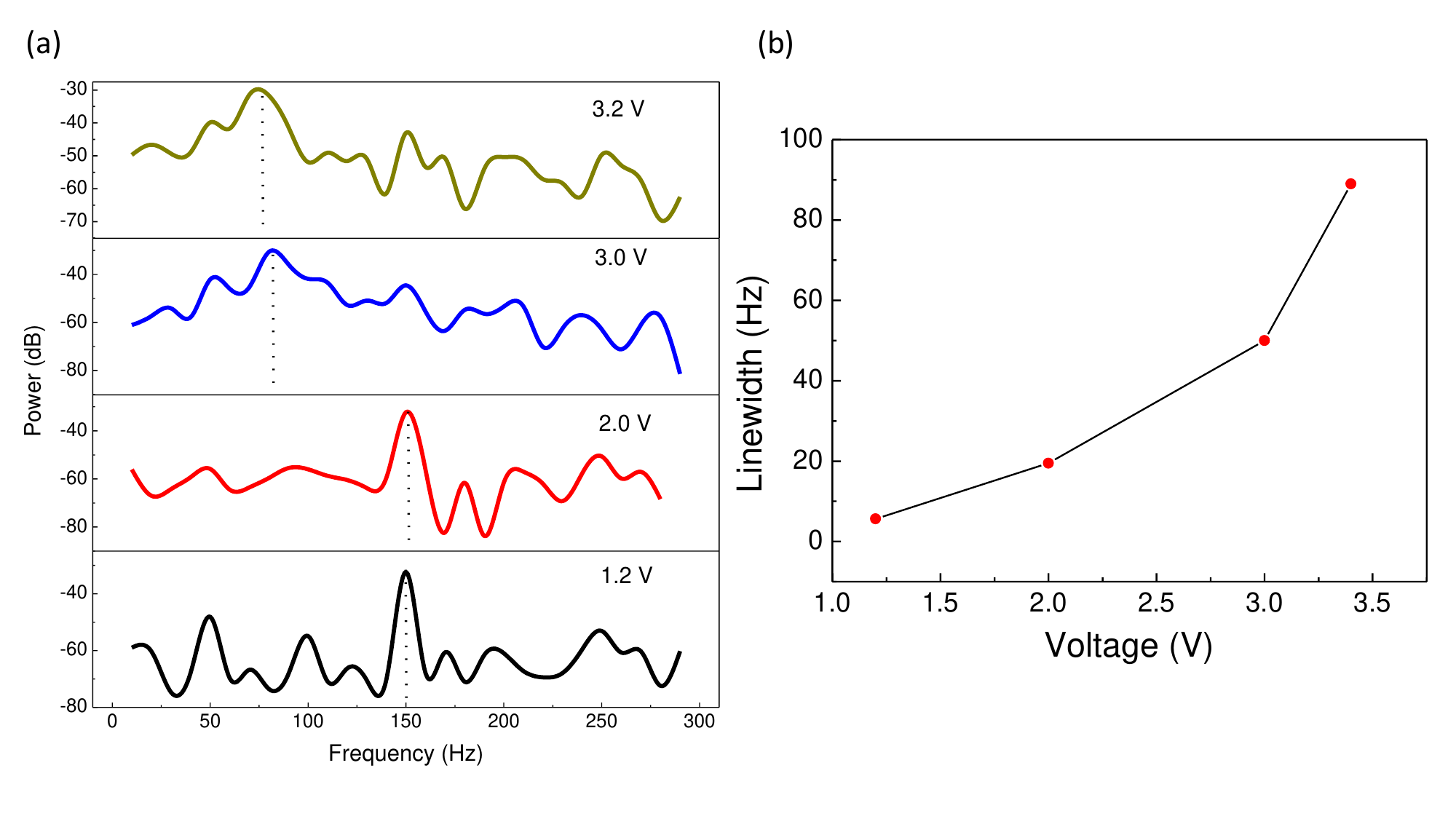}
	\caption{(a) The output power of the oscillation is stable with respect to the applied voltage, though the frequency decreases with increasing the voltage and the linewidth decreases as shown in (b)}
	\label{Figure_S7}
\end{figure*}

\bibliographystyle{MSP}
\bibliography{Bibliography}